\documentclass[vecphys]{svmult}

\usepackage{latexsym}         
\usepackage{makeidx}         
\usepackage{graphicx}        
\usepackage{multicol}        
\usepackage[bottom]{footmisc}

\begin{document}

\title*{White Dwarfs in Globular Clusters}
\author{S. Moehler\inst{1}\and
G. Bono\inst{1,2,3}}
\institute{European Southern Observatory, Karl-Schwarzschild-Str. 2,
85748 Garching, Germany, \texttt{smoehler@eso.org} \and
Dept. of Physics, Univ. of Rome Tor Vergata, via della Ricerca Scientifica 1, 
00133 Rome, Italy, \texttt{bono@roma2.infn.it} 
\and 
INAF-Osservatorio Astronomico di Roma, via Frascati 33, 00040 Monte
Porzio Catone, Italy 
}
%
%
\maketitle

We review empirical and theoretical findings concerning white dwarfs
in Galactic globular clusters. Since their detection is a critical
issue we describe in detail the various efforts to find white dwarfs
in globular clusters. We then outline the advantages of using cluster
white dwarfs to investigate the formation and evolution of white
dwarfs and concentrate on evolutionary channels that appear to be
unique to globular clusters. We also discuss the usefulness of
globular cluster white dwarfs to provide independent information on
the distances and ages of globular clusters, information that is very
important far beyond the immediate field of white dwarf research.
Finally, we mention possible future avenues concerning globular
cluster white dwarfs, like the study of strange quark matter or plasma
neutrinos.

\section{Introduction} \label{GCsec:Intro}
During the last few years white dwarfs have been the topic of several
thorough review papers focused on rather different aspects. The
interested reader is referred to \cite{Hanlie03} and to \cite{Hans04}
for a comprehensive discussion concerning the use of white dwarfs as
stellar tracers of Galactic stellar populations and the physics of
cool white dwarfs. The advanced evolutionary phases and their impact
on the dynamical evolution of open and globular clusters have been
reviewed by \cite{Kalric10}, while \cite{altetal10} provide a
comprehensive discussion of the use of white dwarfs to constrain
stellar and cosmological parameters together with a detailed analysis
of the physical mechanisms driving their evolutionary and pulsation
properties.

The advantages of using cluster white dwarfs over field white dwarfs
in studying the formation, physical properties, and evolution of these
stars come from a number of properties of star clusters.  {\em i) --}
White dwarfs in globular or open clusters are located at the same
distance and have (in general) the same reddening. Moreover and even
more importantly, when moving from hot to cool white dwarfs the
colours of cluster white dwarfs are systematically bluer than those of
main sequence stars (see Fig.~\ref{GCfig:cmdGC_OCen}). This means that
to properly identify cluster white dwarfs we can use the
colour-magnitude diagram instead of a colour-colour plane. Therefore,
the identification of cool cluster white dwarfs is not hampered by the
thorny problem of colour degeneracy with main sequence stars that
affects field white dwarfs (\cite{Hanlie03}).  {\em ii) --} For
cluster white dwarfs we can trace back the evolutionary properties of
the progenitors, since both the chemical composition and the typical
mass at the turn-off of the cluster are well-known.  
While there
is some evidence that the most massive globular clusters
(e.g. $\omega$\,Cen, NGC\,1851, NGC\,2808, NGC\,6388, NGC\,6441)
probably have a more complicated star formation and chemical
enrichment history, most globular clusters are essentially
mono-metallic systems with respect to iron
with a negligible spread in
age. This provides the opportunity to constrain the initial-to-final
mass relation of white dwarfs and to improve the knowledge of the
physical mechanisms governing their final fate (\cite{Hans04};
\cite{Kalirai07b}).  {\em iii) --} According to current evolutionary
predictions the number of white dwarfs in a globular cluster with an
age of 12\,Gyr and a Salpeter-like initial mass function
($\alpha=2.35$) is about a factor of 300 larger than the number of
horizontal branch stars (\cite{Broc99}). This means that the expected
local density of white dwarfs in a globular cluster is several orders
of magnitude larger than the local densities of the halo, thick disk,
and thin disk white dwarf populations, thus allowing us to observe
large homogeneous samples of white dwarfs without the need for
wide-field surveys.

The main drawback for white dwarfs located in globular clusters is
that they are faint objects severely affected by crowding
problems. Photometric observations are well possible with the
Hubble Space Telescope (HST) and profit from software packages that
allow astronomers to obtain precise measurements even in crowded
fields. Spectroscopy, however, poses a different problem: HST is a too
small telescope and lacks the multiplexing capacities that enable
efficient observations in globular clusters. The disentangling of
overlapping spectra in ground based observations is a rather difficult
process, as the contributions from the different stars vary strongly
with wavelength. Early spectroscopic investigations
(e.g. \cite{Moehler00}; \cite{Moehler04}) were hampered by the fact
that their targets were selected from HST photometry, which had
usually been observed rather close to the cluster cores. Good
photometry of the outer regions of globular clusters facilitate
spectroscopic observations enormously, as the problem of background
subtraction is immensely reduced if there are no bright stars close to
the white dwarfs (e.g. \cite{Davisetal2009}).

In a recent investigation \cite{Davies07} called attention to the
evidence that the radial distribution of young white dwarfs in
NGC\,6397 is significantly more extended than the radial distribution
of both older white dwarfs and the most massive main sequence
stars. To account for this peculiar trend they suggest that the white
dwarfs in this cluster do not experience a quiescent birth, but they
receive a natal kick. This scenario has been suggested originally
by \cite{Fellhaueretal2003} and received additional support from the 
theoretical investigations of \cite{Heyl07a,Heyl07b,Heyl07c}. In
particular, he found that asymmetric winds on the asymptotic giant
branch can generate a kick affecting the trajectory of the resulting
white dwarfs. This mechanism would explain why young white dwarfs are
less centrally concentrated than their progenitors. 
The radial distribution of young white dwarfs in $\omega$\,Cen
observed by \cite{Calamidaetal2008b} provides some support for that scenario.

The impact that energetic white dwarfs (i.e. the young white dwarfs
affected by the velocity kick) have on the structural properties of
the host cluster has been investigated by \cite{HeylPenrice2009}.
They found that the white dwarf kicks lose a significant fraction of
their energy in the central region of the cluster. This means that
these objects can either delay the core collapse or
increase the size of the cluster cores. These predictions were soundly
confirmed by \cite{Fregeauetal2009} using a Monte Carlo cluster
evolution code. They found that for globular clusters with velocity
dispersions similar to the kick speed, the white dwarfs' kicks can
delay the phase of core contraction and increase by one order of
magnitude the current ratio between the core and the half-mass
radius.

If this kick effect is
indeed present in other globular clusters as well, it will be very
helpful for the study of white dwarfs in globular clusters, as it
would put the youngest (and therefore brightest) white dwarfs into the
less crowded outer regions.

Beyond the obvious observational complications, we still lack a
detailed understanding of the impact of the high density
environment of globular clusters on the formation and evolution of
cluster white dwarfs (\cite{Monelli05}).

As the exploration of white dwarfs in globular clusters is an
extremely active and rapidly growing field we want to
state here that this review contains information from papers or
preprints (accepted for publication) available by December 2010.

\section{Searches and Detections} \label{GCsec:Search}

As already mentioned above, white dwarfs in globular clusters are not
the most easily observable stars. Their direct detection is
hampered by the severe crowding problems in observations deep enough
to locate white dwarfs, which explains why it took so long
to really identify the first white dwarfs in a globular cluster.

\subsection{Direct Identification} \label{GCsec:SearchDir}

The pioneering ground-based photometric investigations concerning
white dwarfs in globular clusters date back to the late 1980's
(\cite{Orto87} -- $\omega$\,Cen; \cite{Richer88} -- M\,71). However,
detailed investigations of globular cluster white dwarfs were
undertaken only after the refurbishment of the HST with its Wide Field
and Planetary Camera 2 ({\bf WFPC2}).
Several candidate white dwarfs were then soon identified in quite a
few globular clusters. One should keep in mind that the masses
assigned to white dwarf sequences in globular clusters are based on
comparisons with theoretical tracks and therefore strongly depend on the
assumed distance modulus and reddening of the respective cluster. The
clusters are listed by the date of the first discovery paper.

\begin{description}
\item [{\bf NGC\,5139} ($\omega$\,Cen) {\bf -- }] 
This is the most massive Galactic
globular cluster and it is therefore not surprising that the search
for white dwarfs dates back to \cite{Orto87}, who detected two dozen
white dwarf candidates using photometry observed with the Faint Object
Spectrograph and Camera 1 at the ESO 3.6m telescope, and to
\cite{Elson95}, who detected four white dwarf candidates using WFPC2
data. More recently, \cite{Monelli05} detected more than 2,000 white
dwarf candidates in three out of the nine pointings observed with the
Advanced Camera for Surveys ({\bf ACS}) on board the HST, which were located
across the cluster center. Moreover, deep $H_\alpha$ measurements
support the evidence that about 80\% of these cluster white dwarfs are
also $H_\alpha$-bright. In order to account for this empirical
evidence \cite{Monelli05} suggested that a fraction of the white dwarf
candidates might be He-core white dwarfs, and therefore the
aftermath of a violent mass loss event (see
end of Sect.~\ref{GCsec:Helium} for more details).  
Interestingly enough, \cite{Cala08} also noted that candidate
He-core white dwarfs have been identified in stellar clusters
showing evidence for extreme mass loss on the red giant branch, like
the metal-rich old open cluster NGC\,6791 (presence of extreme
horizontal branch stars) and the globular clusters $\omega$\,Cen and
NGC\,2808 (presence of late hot helium flashers,
cf. Sect.~\ref{GCsec:Helium}). Such extreme mass loss could in turn
also yield stars that avoid the He-core flash completely and thus
become He-core white dwarfs.

\begin{figure}
\centering
\includegraphics[width=0.9\textwidth]{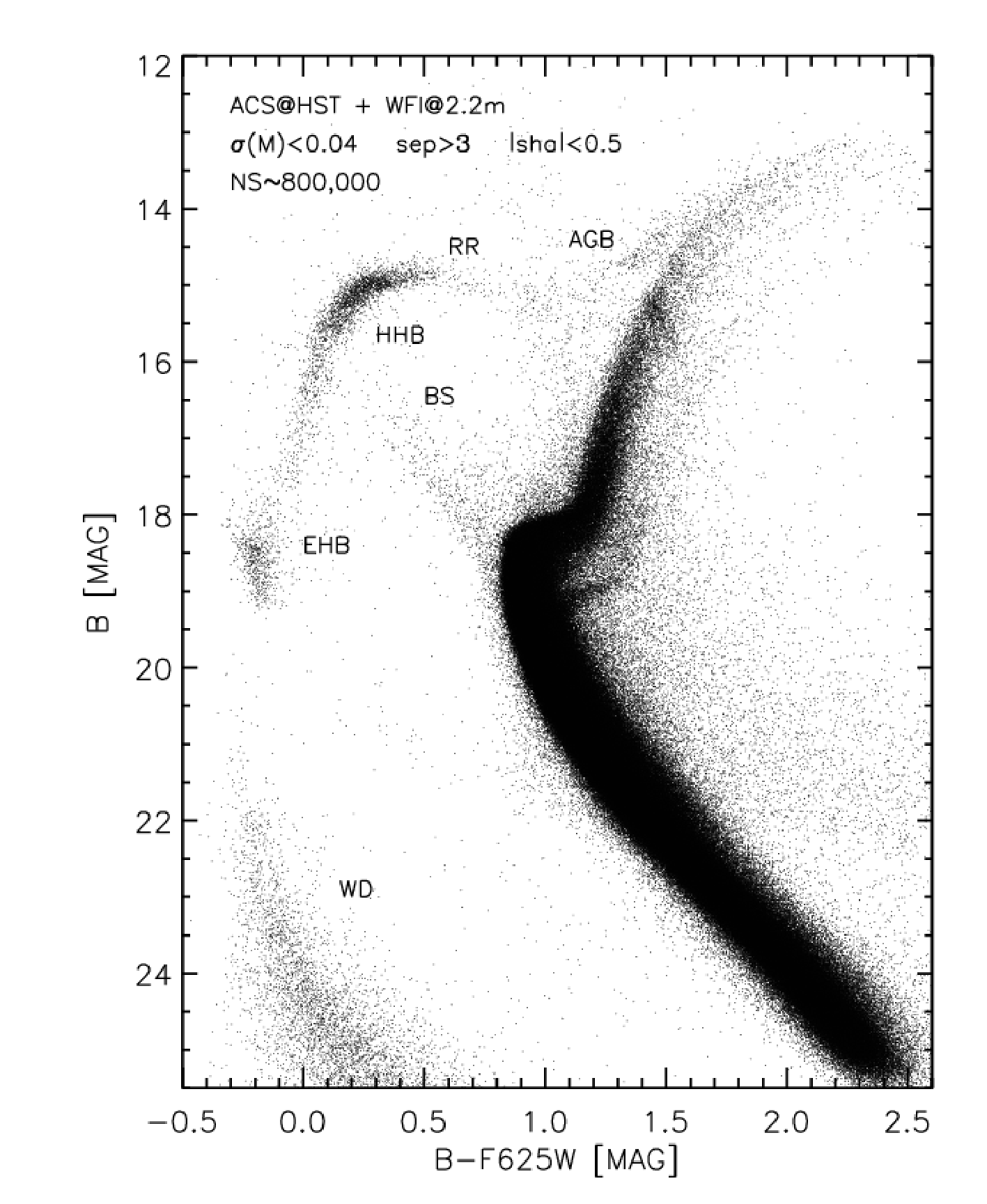}
\caption{ Composite $B, B-F625W$ colour-magnitude diagram of
$\omega$\,Cen based on data collected with the ACS
 and with the Wide Field Imager (WFI)
available at the 2.2m ESO/MPG telescope (\cite{Cast07}; \cite{Cala08}).
The final catalog includes 1.7 million stars. The stars plotted in
this diagram were selected according to {\em sharpness}, {\em
separation} and intrinsic photometric error (see labeled values).
 From top to bottom the labels mark specific evolutionary phases: the
Asymptotic Giant Branch ({\bf AGB}), the RR Lyrae instability strip (RR),
the Hot Horizontal Branch (HHB), the Blue Stragglers (BS), the Extreme
Horizontal Branch (EHB, including hot helium flashers, see
Sec.~\ref{GCsec:Helium}), and the White Dwarf (WD) cooling sequence.
}
\label{GCfig:cmdGC_OCen}       
\end{figure}

\item [{\bf NGC\,6838} (M\,71) {\bf -- }] 
On the basis of deep multi-band data observed
  with the Canada-France-Hawaii Telescope
\cite{Richer88} detected a dozen white dwarfs in this metal-rich
globular cluster.  By adopting an apparent distance modulus of
$(m-M)_V = $ 13.70 they argued that the cluster white dwarf
sequence was either too faint or too blue when compared to DA field
white dwarfs, but in plausible agreement either with DB field white
dwarfs or with a sequence of DA white dwarfs about 0.1\,M$_\odot$
more massive than typical DA white dwarfs in the Galactic disk. 
The small number of objects and the shallow depth of the
colour-magnitude diagram puts these conclusions on somewhat weak
footing and deeper observations would be very helpful. Moreover, they
identified a parallel sequence of blue objects, located between the
white dwarf sequence and the main sequence, which they identified as a
sequence of cluster cataclysmic variables.

\item [{\bf NGC\,6397 -- }] By using deep WFPC2 observations
  \cite{Paresce95a} first detected a candidate white dwarf sequence in
  NGC\,6397, which showed good agreement with theoretical predictions
  for white dwarfs with a mass of 0.5\,M$_\odot$ (assuming an 
  apparent distance modulus of $(m-M)_I = 12.2$) for stars brighter
  than $m_{814} \approx 24.5$. At fainter magnitudes the observed
  numbers exceeded the predicted ones, probably due to contamination
  by, e.g., background galaxies. Independent observations of the same
  cluster with WFPC2 were collected by \cite{Cool96}. They also
  detected the white dwarf sequence, which they fit with models for
  M$_{\rm WD}$ = 0.55$\pm$0.05\,M$_\odot$ with an intrinsic dispersion
  below 0.05\,M$_\odot$ (for a true distance modulus $(m-M)_0 =
  11.9$ and a reddening of $E_{B-V} = 0.18$). This comparison
  illustrates the uncertainties associated with a straightforward
  fitting of observed white dwarf sequences with theoretical models.
  Using deep ACS photometry (down to $m_{F814W} = 28$) of
  NGC\,6397 \cite{Hansen07} detected securely for the first time in a
  globular cluster the blueward turn of the white dwarf sequence
  expected due to collisionally induced absorption ({\bf CIA}) by H$_2$ in
  H-rich white dwarf atmospheres. In their paper they also
  provide a very good discussion of the various parameters affecting a
  fit of the white dwarf cooling curve with theoretical models.

  In 1998 \cite{Cool98} detected seven centrally concentrated
  UV-bright objects in WFPC2 observations together with the bright
  end of the white dwarf cooling sequence. Three of the UV-bright
  objects had been previously identified as cataclysmic
  variables. While a fourth one is also a probable cataclysmic
  variable candidate, the remaining three show no evidence for
  variability (so-called ``Non-Flickerers'') and are good candidates
  for He-core white dwarfs. In 2001 \cite{Taylor01} found three
  additional ``Non-Flickerers'' from WFPC2 observations. These
  ``Non-Flickerers'' form a sequence parallel to the main white dwarf
  sequence, but brighter by about 2 magnitudes in $V_{555}$. The
  fainter ``Non-Flickerers'' ($V_{555} \approx 23$) show strong
  $H_\alpha$ absorption ($H_\alpha-R_{675} > 0.5$). The
  ``Non-Flickerers'' are significantly more centrally concentrated
  than the main sequence stars. Since mass segregation in
  globular clusters causes the more massive stars to sink towards the
  center, this argues for fairly massive companions, if these objects
  are indeed He-core white dwarfs. Due to the observed blue
  broad-band colours, which rule out all but the lowest mass main
  sequence stars (cf. Sect.~\ref{GCsec:wdbin}), the most probable
  companions are white dwarfs.
\item [{\bf NGC\,6121} (M\,4) {\bf -- }] Using WFPC2 data \cite{Richer95} found a
  well defined white dwarf sequence, which is again in agreement with
  a mean mass of the white dwarfs of 0.5\,M$_\odot$ for an 
  assumed distance modulus $(m-M)_V = 12.65$ (note that this is
  close to the best-fit distance of $(m-M)_V$= 12.57 derived by
  \cite{Hansen04}) and an intrinsic dispersion below
  0.05\,M$_\odot$. Here the observed number of white dwarf candidates
  lies below theoretical predictions at fainter magnitudes, possibly
  due to uncertainties in the completeness corrections, to non-DA
  contributions and/or to incorrect
  assumptions concerning the C/O core compositions of the white
  dwarfs.  Subsequent deeper observations collected with the same
  instrument allowed \cite{Richer02} and \cite{Hansen02} to trace the
  white dwarf sequence down to $m_{606W} \approx 30$, which they used
  to derive an age for the cluster (see Sect.~\ref{GCsec:Age}).
\item [{\bf NGC\,104} (47\,Tuc) {\bf -- }] By using the Faint Object Camera on board the
  HST \cite{Paresce95b} found 9 objects in the white dwarf region of
  the colour-magnitude diagram.  In 2001 \cite{Zoccali01} detected the
  white dwarf sequence in WFPC2 observations down to $m_{555} \approx
  27$ and used it to determine the age and the distance of this
  cluster (see Sects.~\ref{GCsec:Age} and \ref{GCsec:Dist},
  respectively).  UV observations with WFPC2 by \cite{Ferraro01} show
  a sequence of 12 hot white dwarf candidates (in agreement with
  theoretical expectations) as well as a large number of objects
  between the blue straggler and the white dwarf
  sequence. Unfortunately, \cite{Zoccali01} and \cite{Ferraro01} were
  published within a few months of each other so that the authors
  could not compare their results. The search for cluster white dwarfs
  in 47\,Tuc has been hampered by the fact that this cluster is
  projected onto the wing of the Small Magellanic Cloud
  (\cite{Zoccali01}).  However, by adopting both far-UV ({\bf FUV})
  images collected with the Space Telescope Imaging Spectrograph ({\bf
  STIS }) on board the HST and F336W images collected with WFPC2
  \cite{kni02} detected a well defined white dwarf sequence in
  the center of the cluster. Thanks to the strong temperature
  sensitivity of the FUV-optical colour they also identified a
  sequence of candidate cataclysmic variables (16) located between the
  white dwarf and the main sequence stars. According to theoretical
  and statistical arguments they estimate that at least a half of
  these objects should belong to the cluster.
\item [{\bf NGC\,6752 -- }] Data collected with WFPC2 allowed 
  \cite{Renzini96} to identify the white dwarf cooling sequence and to
  determine the distance of the globular cluster (see Sect.~\ref{GCsec:Dist}).
\item [{\bf NGC\,2808 -- }] Near-UV ({\bf NUV}) and FUV STIS
  observations of NGC\,2808 (\cite{Dieball05}, see also
  \cite{Brown01}) reveal a population of candidate hot white dwarfs
  (22 objects brighter than $m_{NUV} \approx$ 21) and also a number of
  cataclysmic variable candidates lying between the white dwarf and
  the main sequence in the UV colour-magnitude diagram (about 50
  objects, two of which are variable).

\item [{\bf NGC\,7078} (M\,15){\bf -- }] A clear identification of cluster white dwarfs
 was recently provided by \cite{die07} using deep FUV and NUV ACS
 images.  In particular, they identified 28 white dwarfs and $\approx
 60$ candidates for cataclysmic variables. These findings appear very
 promising, since M\,15 is a very metal-poor ([Fe/H]=$-$2.26,
 \cite{har96}) globular cluster characterized by a very dense core
 ($\rho_c=7\times 10^6 M_\odot pc^{-3}$, \cite{vandb06}) and a very
 high central velocity dispersion \cite{dru96}.  In order to account
 for these features the presence of an intermediate mass black hole
 (\cite{ger02}) or a high mass concentration caused by mass
 segregation of neutron stars and massive white dwarfs (\cite{baum03})
 have been suggested.

\item [{\bf NGC\,6093} (M\,80) {\bf -- }] Two dozens of candidate white dwarfs have been recently
 identified by \cite{Dieballetal2010} in M80 using FUV 
 and NUV ACS images.

\item [{\bf NGC\,3201 -- }] A dozen of bright candidate white dwarfs have also 
 been recently identified by \cite{bonoetal2010} in NGC\,3201 using
 deep V,I-band ACS images at a limiting magnitude of
 V$\approx$I$\approx$24.5 mag. 

\end{description}

In summary, observations of white dwarfs in globular clusters
  indicate that the cooling sequences in general agree with
  theoretical expectations. Unusual populations observed in other
  regions of the colour-magnitude diagram (most notably along the
  horizontal branch), however, also affect the white
  dwarf cooling sequence (e.g. the He-core white dwarf candidates
  in $\omega$\,Cen, see beginning of this section, and NGC\,6791, see
  Sect.~\ref{GCsec:Helium}).

\subsection{Indirect Identification} \label{GCsec:SearchBin}

So far we discussed the identification of white dwarfs in
  globular clusters by looking for their cooling sequence in the
  colour-magnitude diagrams. The detection of white dwarfs in binary
  systems in the clusters requires rather different methods. As the
  section below will show, however, ignoring them would severely bias
  our knowledge about white dwarfs in globular clusters.

  Current theoretical predictions accounting for dynamical and
  evolutionary processes suggest that the binary fraction in dense
  globular clusters, with ages ranging from 10 to 14\,Gyr, is of the
  order of 10\% (\cite{Ivanova05}; \cite{Davies06}; \cite{Hur07};
  \cite{Fre07b}). However, theoretical predictions concerning the
  evolution with time of the binary fraction in globular clusters are
  not very well established yet. Predictions based on full N-body
  simulations and including binary stellar evolution indicate that the
  binary fraction in the cluster core increases with time
  (\cite{Hur07}). On the other hand, predictions based on a simplified
  dynamical model and accounting for binary stellar evolution suggest
  that the binary fraction in clusters decreases with time
  (\cite{Ivanova05}).  The two different approaches to account for the
  fraction of binary stars detected in globular clusters require a
  modest and a large fraction of primordial binaries, and in turn
  different collisional rates.  The reason for this discrepancy is not
  clear, but it has been suggested that the above difference is mainly
  due to differences in the adopted initial conditions and physical
  assumptions (\cite{Fre07a}).  

On the other hand, empirical evidence based on continuous photometric
  monitoring (8.3 days) with WFPC2 of some 46,000 main sequence stars
  in 47\,Tuc indicates that the overall binary frequency is about
  $14\pm4$\% (\cite{Albrow01}).  This estimate accounts for detached
  eclipsing binaries and for contact binaries and might still be a
  lower limit. From the overluminous white dwarfs in M\,4
  \cite{Hansen04} estimate a binary fraction among the white dwarfs of
  0.11$\pm$0.05, in agreement with the percentage of hard binaries
  among field white dwarfs (soft binaries would be destroyed in M\,4).
  By using the colour distribution of main sequence stars, based on
  deep ACS images for 13 low-density globular clusters, \cite{Sol07}
  found that the current minimum fraction of binary stars is of the
  order of 6\% inside the core radius, while the global fraction
  ranges from 10\% to 50\%. According to the above numbers it is not
  surprising that globular clusters should host a broad range of
  exotic objects that show up as faint X-Ray sources. Of those, the
  following have been identified in globular clusters so far: low-mass
  X-Ray binaries (a neutron star accreting from a companion) and their
  progeny, the millisecond pulsars; chromospherically or magnetically
  active binaries (two main sequence stars or a main sequence star and
  an evolved companion); and cataclysmic variables (a white dwarf
  accreting from a low-mass companion). The formation rate of these
  objects in globular clusters is expected to be orders of magnitude
  higher than in the Galactic disk (\cite{Katz75};
  \cite{Clark75}). The advent of the {\em Chandra} X-Ray Observatory
  and of HST provided the opportunity to investigate these objects in
  the central regions of globular clusters (see \cite{Verb06} for an
  extensive review of X-ray sources in globular clusters).  The
  cataclysmic variables appear to be the progeny of either primordial
  binaries or stellar encounters and they have been identified in
  several globular clusters. In particular, $Chandra$ X-ray
  observations in combination with HST photometry identified nine
  cataclysmic variables in NGC\,6397 (\cite{Taylor01}), 22 in 47\,Tuc
  (\cite{Heinke03}; \cite{Heinke05}), and two (possibly three) in M\,4
  (\cite{Bassa04}).  Additional optical counterparts of faint
  Chandra X-ray sources in the post core collapsed globular cluster
  NGC\,6397 have been recently provided by \cite{Cohnetal2010}. They
  identified more than a dozen cataclysmic variables and found that
  they split into two groups: the brighter one in which the optical
  emission is dominated either by the companion or by the accretion
  disk and the fainter one in which the optical emission is dominated
  by the white dwarf. They also found that the former group is more
  centrally concentrated than the latter one and suggested that the
  difference might be caused by an evolutionary process. The
  cataclysmic variables are formed in the very innermost regions of
  the globular cluster by dynamical interaction and are then scattered
  to larger distances as they age. Using FUV STIS spectra
  \cite{Kniggeetal2008} identified four dozens of stellar exotica in
  47\,Tuc. They found that the number of observed cataclysmic
  variables agrees with the predicted one within a factor of 2--3. 

  Optical and FUV data collected with HST and covering long time
  intervals (several years) also provided the opportunity to identify
  outbursting cataclysmic variables in a few globular clusters
  (\cite{shara04}, \cite{shara05}).  This approach is very useful to
  unambiguously identify cluster cataclysmic variables, since the spectroscopic
  follow-up of candidates located in the crowded central regions is
  difficult (\cite{kni03}).

On the basis of these X-ray and optical surveys of a dozen globular
clusters a remarkable correlation has been found between the stellar
encounter rate and the number of X-ray sources in the cores of
globular clusters (\cite{Pooley03}). 

Contrary to expectations, simulations by \cite{Shara06} showed that
the production rate of cataclysmic variables in globular clusters is
comparable to that observed in the field. Due to the high density
environment in globular clusters, however, hard binaries dominate
there, which have a faster evolution.  Moreover, the simulations also
predict a new class of cataclysmic variables that are formed by
exchange interactions, i.e. without a common envelope phase.  Such
cataclysmic variables are formed preferably with donor masses on the
main sequence of more than 0.7\,M$_\odot$ and are short-lived when
compared to classical field cataclysmic variables.  These generally
shorter lifetimes for globular cluster cataclysmic variables imply a
reduction in their expected number, at any given time, by roughly a
factor of three. Another interesting prediction of these simulations
is that the canonical 2--3 hours period gap observed among field
cataclysmic variables should be smeared out in the dynamically formed
globular cluster cataclysmic variables.

Numerical simulations by \cite{Ivanova06} accounting for binary
formation via physical collisions (based on smoothed particle
hydrodynamics) and for the metallicity dependence in the formation and
evolution of close binaries also predict that the formation rate of
cataclysmic variables and AM CVn systems (white dwarf binaries with
one of the two white dwarfs undergoing a Roche lobe overflow) in the
field and over the entire cluster are very similar. There is, however, a larger
variety of formation
channels for binaries in globular clusters than for their field
counterparts: primordial binaries, post-exchange binaries, and
physical collisions between main sequence and red giant stars.

Ultraluminous X-ray sources have X-Ray luminosities well above
the Eddington limit for a neutron star accreting helium and it has
been suggested that these systems might harbor a black hole (BH) X-ray
binary (\cite{Angelinietal2001}; \cite{Brassingtonetal2010}). Such X-ray
sources have been recently identified in a Galactic globular cluster
(NGC\,4472, \cite{Maccaroneetal2007}) and in globular clusters
belonging to external galaxies (\cite{Kimetal2006};
\cite{Irwinetal2010}). Their X-Ray properties suggest that these
systems might consist of a stellar-mass black hole and a compact donor
(i.e. they are black hole--white dwarf binaries).  The formation rates
of such system in globular clusters was investigated by
\cite{Ivanovaetal2010}.  They accounted for binary exchange and
physical collisions and found that the only possibility to form black
hole--white dwarf binaries is via hardening and/or the formation of a
triple system. Moreover, they suggest that in order to explain the empirical
evidence with black hole--white dwarf binaries between 1\% and 10\% of
the black holes present in the core of a globular cluster should
interact with the stellar population located inside the cluster core.

\section{Spectral Types}\label{GCsec:Spec}
Photometric observations alone are generally not sufficient to
distinguish DA from non-DA stars, although H-rich DAs and
He-rich DBs in principle can be distinguished by their photometric
properties alone in the temperature range
10,000\,K\,$\leq$\,T$_{\mathrm{eff}}$\,$\leq$\,15,000\,K (see
\cite{Bergeron95}). Based on this method \cite{Renzini96} classified
two white dwarfs in NGC\,6752 as DBs. However, without a spectroscopic
confirmation, those two stars can also be explained as high-mass DA
white dwarfs, possibly a product of merging. Also the brightest
white dwarf in M\,4 ($V$=22.08) might be a hot (27,000K) DB star
(\cite{Richer97}). In 1999 \cite{Edmonds99} showed a spectrum
observed with the Faint Object Spectrograph ({\bf FOS}) on board the
HST covering the $H_\beta$ region of a ``Non-Flickerer'' in NGC\,6397,
which clearly showed that the spectrum was H-rich.  In 2000
\cite{Moehler00} published the first ground-based spectra of four
white dwarfs in a globular cluster (NGC\,6397) observed with the FOcal
Reducer/low dispersion Spectrograph 1 ({\bf FORS1}) at the
ESO-VLT, which covered the range from 3800\,\AA\ to more than
5000\,\AA. All stars were DA white dwarfs.  Investigating the
ratio between DA and non-DA white dwarfs for effective temperatures
ranging from 10,000 to 14,000 K in NGC\,6397 \cite{Strickleretal2009}
found that only 1, probably 4, out of sample of 126 C/O white dwarfs
are of the non-DA type. Detailed analysis for field white dwarfs
\cite{TremblayBergeron2008a}, however, indicates that the ratio
between the two different samples is 4:1 for stars between 10,000\,K
and 15,000\,K.

Later \cite{Moehler04} also showed FORS1 spectra of five white
dwarfs in NGC\,6752, all of which were once again H-rich.

In 2005 \cite{Kalirai05} showed spectra of 21 white dwarfs in the
young (650\,Myr) open cluster NGC\,2099 ($>$30\% of the known white
dwarfs in that cluster) observed with the Gemini Multi-Object
Spectrograph at the Gemini telescope and with the Low-Resolution
Imaging Spectrometer ({\bf LRIS}) at the Keck telescope, none of which
was He-rich. Cluster members have estimated effective temperatures
between 13,000\,K and 18,000\,K in the observed magnitude
range. Allowing for 4--5 contaminating field stars, one would expect
to find 4 DBs from the 4:1 DA:non-DA ratio observed among field white
dwarfs of similar temperatures (\cite{Salaris01},
\cite{TremblayBergeron2008a}). It was noted by \cite{Beauchamp96} that
the mass distribution of field DB stars lacks the high-mass tail
observed for DA white dwarfs.  Newer observations by \cite{Voss07}, on
the other hand, find no difference in the mass distribution between DB
and DA stars.  Spectroscopic analyses by \cite{Kalirai05a} yield
rather high masses of up to 1.1\,M$_\odot$ for the bright white dwarfs
in NGC\,2099, consistent with the youth of that cluster.  Several
explanations for the possible lack of DB stars among massive white
dwarfs were discussed by \cite{Kalirai05}: \emph{i) --} More massive
white dwarfs have thicker hydrogen layers, which may prevent the
formation of a convection zone mixing helium to the hydrogen
layer. Such a convection zone has been suggested to cause the
appearance of DB stars at about 30,000\,K (\cite{Liebert87};
\cite{Hanlie03})). This explanation, however, has no quantitative
support so far and the currently assumed hydrogen layers are much
thicker than those assumed by \cite{Liebert87}. \emph{ii) --} The
removal efficiency for the hydrogen layer depends on mass. \emph{iii)
--} Binary evolution may result preferably in DA white dwarfs (in
binary systems and also as single stars, \cite{Liebert05}).

Also LRIS spectroscopy of white dwarfs in NGC\,6791 by
\cite{Kalirai07a}, showed only H-rich spectra for all nine
targets with sufficient S/N.

The situation changed recently, when \cite{Williams07} and
\cite{Kalirai07b}, both using LRIS at Keck, detected DB type white dwarfs
in the open clusters NGC\,6633 (0.5\,Gyr) and NGC\,6819 (1.5\,Gyr),
respectively.  

Observations of 19 probable cluster white dwarfs in M\,4 by
\cite{Davisetal2009} show all stars to be H-rich, whereas 4-5
He-rich ones would have been expected (assuming a ratio of
4.2:1). Combining their data with results for other globular and open
clusters (for a total of 140 white dwarfs) the authors derive a probability of
6$\cdot$10$^{-9}$ that the DA:non-DA rate in clusters is the same as
in the field. Due to the large range in mass, age, and metallicity
covered by the sample the authors rule out these parameters as
possible explanations. Suppression of the formation of non-DA stars by
the density in the clusters is ruled as the required density is above
that of most globular clusters and far above any open
cluster. Transforming a non-DA into a DA via accretion is also ruled
out. The finding therefore remains unexplained for now.

\section{White Dwarf Formation and Evolution}\label{GCsec:Evol}

The minimum mass value for carbon ignition in the core is the
so-called M$_{up}$, while M$_{mas}$ defines the minimum mass value to
ignite neon at the center.  Stellar structures with initial
masses smaller than M$_{up}$ produce degenerate C/O cores during the
asymptotic giant branch phase. Current evolutionary prescriptions
(\cite{Bono00}; \cite{Maeder00}; \cite{Girardi02}; \cite{Pietri06})
suggest that these stars finish their evolution either as C/O white
dwarfs (efficient mass loss along the asymptotic giant branch) or with
a carbon deflagration (\cite{Iben83}; \cite{Hil00}) if the mass of 
the C/O core
becomes larger than the Chandrasekhar mass (mild mass loss along the
asymptotic giant branch).  Stellar structures with initial masses
larger than M$_{mas}$ experience all the nuclear burning stages and
finish their evolution as iron core collapse supernovae (\cite{Woo02};
\cite{Lim03}).  Stellar structures with initial masses between
M$_{up}$ and M$_{mas}$, the so-called super-AGB stars, start to ignite
carbon in a partially degenerate off-center shell. After a few shell
flashes these structures eventually start to burn carbon in the center
and finish their evolution either as massive O/Ne white dwarfs
(\cite{Nomoto84}; \cite{Tornambe84}; \cite{Gilpons05}; \cite{Siess07})
or as electron-capture supernovae where the core collapse is triggered
by electron captures before the neon ignition
(\cite{Wan03}). Evolutionary calculations indicate that the value of
M$_{up}$ strongly depends on the assumptions adopted to deal with
semi-convection and convective overshooting. By using three
evolutionary codes with different mixing algorithms, \cite{Poela08}
found that, at fixed metal abundance, the value of $M_{up}$ ranges
from 7.5 to 9.0\,M$_\odot$.

White dwarfs in globular clusters thus have a broad range of
progenitor masses, but they are currently produced by progenitors with
stellar masses of about 0.8-1.0\,M$_\odot$ (\cite{Richer00}). Due to
their long cooling times massive white dwarfs in globular clusters
will generally be extremely faint.  However, massive cluster white
dwarfs may currently evolve from blue stragglers or from the merging
of close white dwarf binaries (see, e.g., \cite{Marsh95}). The general
evolutionary properties of white dwarfs have been investigated in many
papers (see, e.g., \cite{Koester90}; \cite{Dantona90};
\cite{Benvenuto97}; \cite{Althaus98}; \cite{Serenelli02};
\cite{Fontaine05}; and references therein).  New cooling
sequences for white dwarfs by \cite{Salarisetal2010} take into account
accurate boundary conditions, which are based on model atmospheres and on C/O
chemical abundance profiles (based in turn on detailed evolutionary
calculations from the BASTI data base).  The authors also analysed in detail
the impact of the input physics (mixing during central H and He
burning phases, number of thermal pulses, progenitor metallicity,
$^{12}$C($\alpha,\gamma$)$^{16}$O reaction rate) on the accuracy of
the predicted cooling times. They found that the leading factors are
the treatment of the convection during the final phases of He burning
and the $^{12}$C($\alpha$,$\gamma$)$^{16}$O reaction rate.

The physical mechanisms driving the crystallization of white dwarfs
has been investigated recently by \cite{winget09}. They compare
theoretical predictions 
with the colour-magnitude diagram and luminosity function of cluster
white dwarfs in NGC\,6397 (\cite{Hansen07};
\cite{richeretal08}). They found that the crystallization is a
first-order phase transition and releases latent heat during this
process as originally suggested by \cite{vanhorn68}. They also found
that the melting temperature of white dwarf cores approaching the peak
of the luminosity function is close to that of pure carbon. Moreover,
they mention that the comparison between predicted and empirical white
dwarfs cooling sequences can be adopted to constrain not only the
ratio between coulomb and thermal energy close to onset of
crystallization, but also the C/O ratio in the center and
the energy released by crystallization due to the phase separation of
carbon and oxygen.

Using molecular dynamics simulations accounting for both liquid and
solid phases \cite{horowitz10} investigated the phase diagram for C/O
white dwarfs.  Based on the results of
\cite{winget09} concerning the melting temperature of white dwarf
cores they predict the abundance of oxygen in these structures to be
less than 60\%. This evidence together with predictions concerning the
treatment of convection in evolutionary models (\cite{salaris97}) can
be adopted to constrain the effective astrophysical S factor of the
$^{12}$C($\alpha$,$\gamma$)$^{16}$O nuclear reaction.  

In the following
subsections we briefly discuss the formation of {\em low-mass white dwarfs}
from binary evolution and otherwise.

\subsection{Binary Evolution} \label{GCsec:wdbin}

Empirical evidence based on field white dwarfs, for which
spectroscopic mass estimates are available, indicates that only a
fraction of the order of 10\% has masses below 0.45\,M$_\odot$,
which is the limiting mass required to remove the electronic
degeneracy and to ignite He-core burning via the He-core flash
(\cite{Cast06}). These He-core white dwarfs are generally
considered to be the aftermath of binary star evolution, since single
stars with such low masses are not expected to evolve to the
white dwarf cooling sequence within a Hubble time. These objects
underwent an episode of extreme mass loss, possibly caused by stellar
encounters or by evolution in compact binaries, while they
approach the tip of the red giant branch ({\bf RGB}, \cite{Kipp67}).

In 2003 \cite{Hansen03} published a study on the effects of binary
evolution on the white dwarfs in globular clusters. Binary formation
by exchange interactions between hard binaries and single stars in
globular clusters promotes the production of He-core white dwarfs
as it increases the average secondary mass in binaries containing C/O
white dwarfs or neutron stars (thereby increasing the probability of
mass transfer episodes). In addition the final He-core white dwarf
will be removed preferably from these systems during ensuing exchange
interactions. Tidal captures followed by mergers or direct stellar
collisions involving red giants result in binaries with a red giant
core (= proto He-core white dwarf) and a more massive companion,
which accreted most of the red giant's envelope (unless that was
ejected). Low-mass He-core white dwarfs are, at fixed effective 
temperature, brighter than typical C/O-core white dwarfs and the 
cooling times of the former objects are significantly longer than 
the latter ones. 
This means that even a small fraction of He-core white
dwarfs significantly increases the number of observed bright
white dwarfs in globular clusters (\cite{Cast94}; \cite{Cast06}, see
also Sect.~\ref{GCsec:SearchDir} for $\omega$\,Cen and
Sect~\ref{GCsec:Helium} for NGC\,6791).

The globular cluster NGC\,6397 shows a centrally concentrated
population of UV-bright stars (\cite{Cool98}; \cite{Edmonds99};
\cite{Taylor01}; see also Sect.~\ref{GCsec:SearchBin}). Spectroscopic
follow-up collected with the FOS by \cite{Edmonds99} indicates that
the observed non-variable UV bright star is a binary system consistent
with a He-core white dwarf and an unseen massive companion
being either a neutron star or a massive white dwarf. This was the
first He-core white dwarf identified in a globular cluster.
Detailed investigations by \cite{Hansen03} assuming various values for
the thickness of hydrogen layers and the companion masses (neutron
stars, C/O white dwarfs) suggest that the ``Non-Flickerers'' in
NGC\,6397 are young objects with C/O white dwarf companions, whereas
neutron star companions do not provide self-consistent solutions. This
is in agreement with the lack of X-ray emission from
``Non-Flickerers'' observed by \cite{Grindlay02}.

Using ACS multi-band photometry of NGC\,6397 \cite{Strickleretal2009}
found two dozen of hot objects forming a well defined sequence
parallel to the canonical sequence of C/O-core white dwarfs, which are
best explained by He-core white dwarfs with masses of 0.2--0.3
M$_\odot$. Moreover, these objects also show red H$_\alpha$-R colors
suggesting that they have strong H$_\alpha$ absorption lines, which
excludes them from the sample of cataclysmic variables already
identified in this globular cluster.  These objects can only be
produced by violent mass loss events (e.g. during binary evolution),
which is also in agreement with their radial distribution being more
centrally peaked than that of C/O white dwarfs and main sequence
stars, but similar to that of the Blue Stragglers. The authors
performed a detailed analysis concerning the nature and the mass of
their companions and found that the companions cannot be main sequence
stars, but have to be heavy C/O white dwarfs.

The position of the second He-Core white dwarf in 47\,Tuc (identified
from FUV spectroscopy by \cite{Kniggeetal2008}) in the
colour-magnitude diagram agrees
quite well with that of candidate He-core white dwarfs
recently identified in $\omega$\,Cen (\cite{Cala08}) and in NGC\,6397
(\cite{Strickleretal2009}). The effective temperature
($\sim$21,000\,K) and the radius (R$\sim$0.05 R$_\odot$) derived
for this object agree quite well with those found by
\cite{Edmonds99} for the He-core white dwarfs in NGC\,6397.  

In an exhaustive review \cite{Ben06} discussed the theoretical 
models of globular cluster evolution and the dynamical interaction between 
single and binary stars. The author also provides a detailed analysis 
concerning the formation and evolution of binary systems including 
one or more compact objects. The observational techniques currently 
adopted to detect and characterize these exotic objects are also reviewed,
including the possibility to detect gravitational radiation from the 
relativistic binaries in globular clusters with LISA. 

\subsection{Exotic Cases}\label{GCsec:Helium}
A well defined white dwarf sequence in the old open cluster NGC\,6791
was observed by \cite{Bedin05} with ACS. The fit with C/O
white dwarf models suggested an age of 2.4\,Gyr, in stark contrast
to the age of 8\,Gyr derived from the main sequence turnoff of this
cluster. The authors ruled out an incorrect distance modulus and
incorrect assumptions on the mass of the hydrogen layer or the C/O
ratio in the cores of the white dwarfs as possible
explanations. He-core white dwarfs from binary evolution also
cannot explain the observed white dwarf sequence, since such white dwarfs
would have low masses and would therefore be too bright.

Prompted by these findings \cite{Hansen05} explored further possible
explanations and found that two effects can contribute to explain
bright white dwarfs in NGC\,6791: {\em i) --} Retardation of cooling
by $^{22}$Ne sedimentation (\cite{Bildsten01}; \cite{Deloye02}), which
would predict the peak of the white dwarf cooling sequence for C/O
white dwarfs at $m_{606W}$$\approx$28--29 (instead of
$m_{606W}$$\approx$29--30 from canonical models). {\em ii) --}
Production of \emph{massive} ($>$0.4\,M$_\odot$) He-core white
dwarfs due to extreme mass loss on the RGB.  The idea of
increased mass loss is supported by the existence of extreme
horizontal branch stars in NGC\,6791
(\cite{Kaluzny92,Liebert94,Landsman98}). This second scenario can
explain the peak in the luminosity function at $m_{606W}$$\approx$27.5
for a cluster age of 8\,Gyr, assuming that more than 50\% of all low
mass stars with main sequence masses above 1.6\,M$_\odot$ become
He-core white dwarfs.  Analyses of LRIS spectra of bright
($V$$\approx$22\ldots24) white dwarfs in NGC\,6791 by
\cite{Kalirai07a} yielded masses below 0.46\,M$_\odot$ (threshold for
core helium burning) for six out of nine probable cluster members,
with the remaining three having masses of 0.47\,M$_\odot$,
0.48\,M$_\odot$, and 0.53\,M$_\odot$, supporting the high mass loss
scenario outlined by \cite{Hansen05}. Spitzer observations by
\cite{vanloon08}, however, provide no evidence for enhanced mass
loss.

Deeper observations by \cite{beki08} show a problem
with the identification of the brightest peak in the white dwarf
luminosity function as He-core white dwarfs. In the new data the
bright peak shows an extension to the blue, which is supposed to be
caused by the more massive and older white dwarfs, which have smaller
radii. In the case of He-core white dwarfs, however, this would
require white dwarfs with masses above 0.5\,M$_\odot$, i.e. above the
minimum mass for He-core burning. The new data show also a second,
fainter peak for which a fit with canonical C/O white dwarfs gives an age
of 6\,Gyr. A possible solution to the puzzling difference with the turn-off
age (8\,Gyr, see their Fig. 10) would be the rotation of white dwarf
progenitors. Rotation increases the minimum mass for the He-core
flash by up to 0.15\,M$_\odot$ \cite{megr76} while at the same time
increasing the mass loss on the RGB.  Current data do not
allow us to verify this scenario.

The brighter peak of the white dwarf luminosity function in NGC\,6791 at
  m$_{F606W} \approx$ 
27.45 might be explained if 34\% of the observed white dwarf were in
white dwarf+white dwarf binaries (\cite{Bedinetal2008}). To achieve
this a binary fraction of 50\% among the main sequence is required,
similar to what has been derived for M67 and consistent with the large
number of interacting binaries observed in the field of NGC\,6791.

Other old open clusters do not show such a discrepancy between
the ages derived from the main sequence and from the white dwarf
cooling sequence (NGC\,2158 \cite{Bedinetal2010}, M\,67
\cite{Bellinietal2010}).

Motivated by the recent discoveries of probable 
He-core white dwarfs in open
clusters (M\,67, \cite{landsman97}; NGC\,6791, \cite{Kalirai07a},
\cite{vanloon08}, \cite{beki08} and in several globular clusters
(NGC\,6397, 47\,Tuc, $\omega$\,Cen), \cite{althausetal2009} performed an
exhaustive theoretical investigation of the evolutionary properties of
He-core white dwarfs with metal-rich progenitors. The key advantage of
their approach is to account in a self-consistent theoretical
framework for gravitational settling, chemical diffusion
and residual nuclear burning. Moreover, they also use LTE atmosphere
models explicitly including Ly$_\alpha$-quasi molecular opacity to
predict the white dwarf colors.  They
found that chemical diffusion at the base of the H-rich
envelope affects the residual nuclear burning during the advanced
phases of white dwarf evolution.  In particular, white dwarf cooling sequences
accounting for diffusion are mainly governed by the thermal content of the
ions, while white dwarf cooling sequences neglecting
diffusion show spuriously longer cooling times (several Gyr) due to
residual hydrogen burning.

The aforementioned grid of white dwarfs cooling sequences was extended
to more metal-poor progenitors, namely Z=0.01 and Z=0.001, by
\cite{renedo10}. The cooling tracks have been followed to low
effective temperatures (2500\,K) and provide a very useful theoretical
framework for white dwarfs in old stellar systems.

The same group in a subsequent investigation (\cite{althausetal2010})
addressed the energy released during the white dwarf cooling sequence
with metal-rich progenitors by the processes of $^{22}$Ne
sedimentation and C/O phase separation upon crystallization.
They found that the former process strongly delays the cooling rate at
moderate luminosities, while the latter does so at low luminosities.  They
also investigated the impact of current uncertainties on $^{22}$Ne
diffusion coefficients on the cooling ages and found that their new
white dwarf models solve the age discrepancy between the white dwarfs
and the main sequence turn-off in NGC\,6791.

A few massive globular clusters (e.g. $\omega$\,Cen, NGC\,2808,
NGC\,6441) show a population of very hot subluminous horizontal branch
stars, the so-called blue hook stars (\cite{DCruz96}; \cite{Brown01};
\cite{Moehler02}; \cite{Moehler04a}; \cite{Busso07};
\cite{Caloi07}). The formation and evolution of these objects prompted
a lively debate in the recent literature. In order to account for
their photometric and spectroscopic properties the so-called ``hot
helium-flasher'' scenario has been suggested (\cite{Cast93};
\cite{DCruz96}; \cite{Brown01}; \cite{Cast06}). In this theoretical
framework red giant stars experience a violent mass loss event, which
decreases the total mass of these structures to below the limit for
central helium ignition before approaching the tip of the red giant
branch. The red giant stars with a mass slightly larger than this
limit undergo a He-core flash at high temperatures either during their
approach to the He-core white dwarf cooling sequence (early hot helium
flasher, \cite{Cast93}; \cite{DCruz96}) or along this sequence (late
hot helium flasher, \cite{Brown01}).  Several evolutionary sequences
for the hot-flasher scenario covering a broad range in heavy element
abundances and physical assumptions are provided by
\cite{MillerBertolamietal2008}. They solve simultaneously for mixing
and nuclear burning and account for convective transport either via a
diffusive equation or via a mixing length approach. They also
investigate the impact of chemical gradients and extra mixing at the
edges of the convective regions, and in particular, the interplay
between diffusion and mass loss.  In particular, they found that
element diffusion during the early He-core burning can transform a
He-rich into a He-deficient atmosphere.

In 2006 \cite{Cast06} studied the consequences of this scenario
for the white dwarf population in globular clusters. A globular
cluster with a significant number of late hot helium-flashers will
also produce a significant fraction of He-core white dwarfs (i.e. the
red giant stars with a mass smaller than the limit for central helium
ignition). At luminosities L$>$0.1\,$L_\odot$ He-core white dwarfs
have a life time comparable to typical extreme horizontal branch
stars. By assuming that a fraction of 20\% of the stars in a globular
cluster do not evolve through a He-core burning phase one expects
twice as many white dwarfs with L$>$0.1\,$L_\odot$ than in the
canonical case (all stars undergoing central helium burning).  This
scenario is supported by \cite{Sand07}, who found a deficit of the
order of 20\% in the number of bright RGB
stars in the globular cluster NGC\,2808 from ACS and WFPC2
photometry. A few dozens of white dwarfs have already been identified
in NGC\,2808 by \cite{Dieball05}, but their limiting magnitude is too
shallow to constrain the impact of the ``missing giants''.

The analysis of FUV--NUV colour-magnitude diagrams of six
massive globular clusters by \cite{Brownetal2010} shows subluminous
stars at the hottest part of the horizontal branch for all
clusters. Normal evolution at standard helium abundance (Y=0.23) can
explain only the 'normal' blue and extreme horizontal branch
stars. Normal evolution at enhanced helium abundance cannot explain
the colour-magnitude diagrams without violating distance and
reddening constraints from independent observations. Flash-mixed
models for both helium abundances reproduce the luminosity range of
the blue hook stars, but not the colour range (especially towards the
red). Some clusters show almost no 'normal' extreme horizontal branch
stars, but only blue horizontal branch and blue hook stars. Apparently
a minimum globular cluster mass is needed to have blue hook stars, but
the cause for that threshold remains unclear.

Using optical (B), NUV and FUV ACS data of the center of M\,15
\cite{Haubergetal2010} study the hot stars in this metal-poor,
core-collapsed globular cluster. The colour-magnitude diagram suggests
a small population of blue hook stars and He-core white dwarfs. It
also shows objects between the blue hook and the He-core white dwarf
region, although any canonical tracks pass through very quickly
through this part of the colour-magnitude diagram ('bright blue
gap'). These objects cluster around the tracks for very young He-core
white dwarfs with thick H envelopes in the colour-magnitude diagram
and are concentrated strongly towards the cluster center, suggesting
binarity. They might thus either be very young He-core white dwarfs or
binaries with currently active mass transfer (one X-ray source shows
similar colours). Very approximate estimates of RGB collision rates in
the cluster center suggest that those may be high enough to explain
the production rates for the He-core white dwarfs. Mass transfer in
close binaries cannot be ruled out, either. The formation rate derived
for the C/O white dwarfs, however, is rather low compared to the rates
at which stars evolve off the main sequence and the horizontal
branch. The objects between the main sequence and the white dwarfs in
FUV vs FUV--NUV show up on the main sequence in B, NUV--B. This might
be explained by a population of detached white dwarf--main sequence
binaries. Alternatively, they could be magnetic cataclysmic variables
with truncated or absent accretion disks.  

The existence of subpopulations enriched in helium
(\cite{Norris04,Lee05,Dantona05,Dantona07}), which has been suggested
to explain the multiple main sequences found in $\omega$\,Cen
(\cite{bedin04}) and NGC\,2808 (\cite{piotto07}) also has consequences
for the white dwarfs in these clusters. Recent dynamical simulations
of a globular cluster, where the first generation of stars enriches
the second generation with helium, have been performed by
\cite{Downing07}. In the case of a top-heavy initial mass function for
the first generation a large number of massive white dwarfs is
predicted compared to the results for a Salpeter initial mass
function. Accurate number ratios of white dwarfs vs. main sequence
stars would allow the authors to verify their models.

Using deep multi-band photometry of $\omega$\,Cen observed with the
ACS and the Wide Field Imager at the ESO/MPG-2.2m telescope
\cite{Cast07} found that empirical star counts of horizontal branch
stars are on average larger (30\%--40\%) than predicted by canonical
models.  The possible occurrence of helium-enhanced stars cannot
account for this excess. The authors suggested that the excess of
horizontal branch stars might be due to ``hot helium-flashers''. This
working hypothesis implies that $\omega$\,Cen should also host a
sizable fraction of candidate He-core white dwarfs
(cf. Sect.~\ref{GCsec:SearchDir}).  The white dwarf cooling
sequence of $\omega$\,Cen was recently investigated by \cite{Cala08} using
a mosaic of ACS images located across the center of the
cluster. They identified more than 6,500 white dwarf candidates and the
observed ratio between white dwarfs and main sequence stars is a
factor of two larger than the ratio between the cooling time of
C/O-core white dwarfs and main sequence lifetime. The possible
occurrence of He-enhanced subpopulations does not solve the
discrepancy, since an increase of the He content from 0.25 to 0.42
causes an increase in main sequence lifetime by only 15\%. The
possible occurrence of He-core white dwarfs might explain the observed
discrepancy, since the cooling time of these structures is slower than
for canonical C/O-core white dwarfs. Plain physical arguments indicate
that the fraction of He-core white dwarfs necessary to explain the
excess of white dwarfs ranges from 15\% to 80\% depending on their
mean mass. These preliminary evidence --- if supported by independent
photometric and spectroscopic investigations --- would imply that the
fraction of He-core white dwarfs in at least some globular clusters might be
significantly higher (at least a factor of five) than among field
white dwarfs.

The possible occurrence of a significant fraction of He-core white
dwarfs in $\omega$\,Cen was independently supported by
\cite{Cassisietal2009}. They performed a detailed comparison between
extreme horizontal branch stars in the so-called {\em ``blue clump''}
region and evolutionary prescriptions and found that these stars can
be explained as a mixture of hot helium flashers, stars with a
He-enriched composition and stars with canonical He content. By
comparing observed star ratios of H and He-burning phases with the
ratio of evolutionary lifetimes the same authors found that at least
15\% of extreme horizontal branch stars are missing in the
colour-magnitude diagram. This evidence further supports the working
hypothesis suggested by \cite{Cala08} that a non negligible fraction
of bright RGB stars end up their evolution as He-core white dwarfs.

\section{Astrophysical Use of White Dwarfs in Globular Clusters}\label{GCsec:Age_Dist}

Globular cluster white dwarfs play a crucial role not only to
validate current evolutionary predictions, but also as a diagnostic to
constrain the ages and distances of globular clusters.

\subsection{Cluster Distance Determinations}\label{GCsec:Dist}

The use of the white dwarf sequence as a standard candle for
determining the distance to nearby globular clusters was suggested by
\cite{Renzini88}.  The approach is similar to the traditional main
sequence fitting procedure using local subdwarfs with known
trigonometric parallaxes. The white dwarf sequence of the cluster is
compared to a sequence constructed from local white dwarfs with
accurate trigonometric parallaxes. While it may seem strange to use
the faintest objects in a globular cluster to derive its distance,
white dwarfs offer some advantages as standard candles when compared
to main sequence stars:
\emph{i) --} They come in just two varieties - either H-rich
(DA) or He-rich (DB) -- {\em independent of their original
metallicity} and, in both cases, their atmospheres are virtually free
of metals.
So, unlike in the case of main sequence fitting, one does
not have to find local calibrators with the same metallicity as the
globular clusters.  \emph{ii) --} White dwarfs are locally much more
numerous than metal-poor main sequence stars and thus make it possible
to define a better reference sample.

However, the method has its own specific problems, which are discussed
in great detail in \cite{Zoccali01} and in \cite{Salaris01}. Most of the
discussion below is taken from these excellent papers. Indeed, the
location of the white dwarf cooling sequence depends on:

\begin{itemize}
\item \emph{The white dwarf mass}\\ On theoretical grounds, given the
observed maximum luminosity reached on the asymptotic giant branch
(AGB), the mass of currently forming white dwarfs in globular clusters
should be $0.53\pm 0.02$\,M$_\odot$\ (\cite{Renzini88};
\cite{Renzini96}).  Recent homogeneous evolutionary computations
(\cite{PradaMoroni07}) from the pre-main sequence to the tip of the
asymptotic giant branch suggest that,
depending on the metallicity, progenitors with a stellar mass of
0.8\,M$_\odot$ form white dwarfs with masses ranging from
0.55\,M$_\odot$ (Z=0.02) to 0.575\,M$_\odot$ (Z=0.0001).
Unfortunately, there are no local white dwarfs in this mass range with
directly determined masses (i.e. without using a mass-radius
relationship).  There is, however, a handful of local white dwarfs
with spectroscopically determined masses near this value (cf. Table 1
in \cite{Zoccali01}), which allows the construction of a
semi-empirical cooling sequence for M$_{\rm WD}$=0.53\,M$_\odot$, once
relatively small mass-dependent corrections are applied to each local
white dwarf.

However, there are also some systematic differences between clusters
(\cite{Salaris01}): At a given metallicity some globular clusters
(e.g.\ NGC\,6752) possess very blue horizontal branches with
horizontal branch star masses as low as 0.50\,M$_\odot$. Such extreme
horizontal branch stars evolve directly to low-mass C/O-core white
dwarfs (bypassing the asymptotic giant branch, therefore also called
AGB-manqu\'e stars, \cite{Greggio90}) and shift the mean white
dwarf mass closer to 0.51\,M$_\odot$. Other clusters show only very
red horizontal branch stars, which will evolve along the asymptotic
giant branch (thermal pulsing AGB) and form preferably white dwarfs
with masses of about 0.55\,M$_\odot$, depending on the mass-loss
efficiency along the asymptotic giant branch.

Also other channels may exist to produce white dwarfs with masses
above or below the cluster mean, as described in
Sect.~\ref{GCsec:wdbin}.  The spectroscopic determination of the white
dwarf masses in a globular cluster was first attempted by
\cite{Moehler00} for white dwarfs in NGC\,6397. However, the low S/N
ratio of the spectra of these very faint stars did not allow them to
determine the mass with sufficient accuracy. Multi-colour photometry
of white dwarfs in NGC\,6752 in combination with low-resolution
spectra allowed \cite{Moehler04} to estimate a most probable mass of
0.53\,M$_\odot$ for the bright white dwarfs in this cluster (assuming
a true distance modulus of $(m-M)_0$ = 13.20 and a hydrogen layer
mass of 10$^{-4}$M$_\odot$).

Using spectroscopy of six white dwarf members of M\,4 to fit
effective temperatures and surface gravities \cite{Kaliraietal2009}
derive masses via the mass-radius relation. The masses are adjusted by
+0.034\,M$_\odot$ as suggested by the model spectra of
\cite{TremblayBergeron2008b} and yield an average mass of
0.53$\pm$0.01\,M$_\odot$ in good agreement with theoretical
expectations.

\item \emph{The white dwarf envelope mass}\\ In the case of DA white
dwarfs the cooling sequence location depends also on the mass of the
residual H-rich envelope. The thickness of the envelope is
important for the energy loss rate of the white dwarf cores and also
affects the white dwarf radius at a given temperature
(\cite{Salaris01}). For ranges in envelope mass from
10$^{-4}$M$_\odot$ to 10$^{-6}$M$_\odot$ the cooling sequences for DA
white dwarfs differ by about 0.07 in $M_V$, whereas non-DA white
dwarfs show no difference.  This also affects {\it spectroscopically}
derived masses (see above), with the resulting mass being about
0.04\,M$_\odot$\ higher when using the {\it evolutionary} envelope
mass ($10^{-4}$M$_\odot$, \cite{Fontaine97}) as opposed to
virtually zero envelope mass. This mass uncertainty corresponds to an
uncertainty of 0.1\,mag in the distance modulus and 1--1.5\,Gyr in the
age derived from the main sequence turnoff.

\item \emph{Spectral type}\\ DB stars are fainter than DA stars at a
given colour, with the offset depending on the filter combination
(i.e. the offset is greater in $V$ vs. $B-V$ than in $I$
vs. $V-I$). In the field we find a ratio of DA:non-DA of about 4:1
(\cite{Salaris01}, \cite{TremblayBergeron2008a}), while the number for
globular clusters appears to be much lower
(cf. Sect.~\ref{GCsec:Spec}). As non-DAs have colour-magnitude
relations different from DAs (DA white dwarfs are, at fixed
luminosity, systematically cooler than DBs) their undetected presence
in the white dwarf cooling sequence can bias the distance
determination. Since the cooling sequences are well separated (e.g. in
$B-V$) this effect can be accounted for by a prudent choice
of filters. However, the possible occurrence of He-core white
dwarfs makes the photometric identification of DA and DB white dwarfs
a risky approach, since He-core white dwarfs, at fixed luminosity,
are also systematically cooler than DB C/O core white dwarfs.

\item \emph{Chemical composition of the core (varying C/O profiles,
He-core white dwarfs)}\\ While the structure of the C/O-core is
not well known, even large changes do not significantly affect the
position of the cooling sequence (\cite{Salaris01}). He-core white
dwarfs are, at fixed mass and effective temperature, brighter than
C/O-core white dwarfs, but their cooling sequences can overlap in the
$M_V, V-I$ plane.

\item \emph{Initial-to-final mass relation}\\ For the initial-to-final
mass relation assumed by \cite{Salaris01} white dwarfs with progenitor
masses below 2.5\,M$_\odot$ have a mass of 0.54\,M$_\odot$, which
increases to 1.0\,M$_\odot$ for progenitor masses of
7\,M$_\odot$. 
Regardless of the assumed initial-to-final
mass relation the masses of the white dwarfs are constant for the
temperature range between 10,000\,K and 20,000\,K (although the actual
value varies with the adopted initial-to-final mass relation). Also
the progenitor masses are basically constant for this temperature
range, since the white dwarf evolution is very fast when compared to
the cluster ages. This is not true for the {\em field} stars, where
due to extended star formation white dwarfs of different ages and
masses can occupy a given temperature range.

Using masses for white dwarfs in M\,4 as well as for white
dwarfs in various open clusters \cite{Kaliraietal2009} derive a linear
initial-final-mass relation with no dependency on metallicity. 

\item \emph{Reddening}\\ An uncertainty of 0.01 mag
  in $E_{B-V}$ yields an uncertainty of 0.055 mag in the true distance
  modulus (\cite{Salaris01}).
\end{itemize}

The first determination of the distance to a globular cluster using
its white dwarf sequence was provided by \cite{Renzini96} using WFPC2
data. Their distance modulus for NGC\,6752 of $(m-M)_0 =
$13.05$\pm$0.1 was a bit lower than, but consistent with, the ones
obtained from main sequence fitting (13.12$\div$13.23, see
\cite{Moehler04} for a discussion). In 2001 \cite{Zoccali01} repeated
this experiment for 47\,Tuc, again using WFPC2. The brightnesses of
the local white dwarfs were corrected to a mass of 0.53\,M$_\odot$
assuming a hydrogen envelope mass of 10$^{-4}$M$_\odot$. They obtained
a true distance modulus of $(m-M)_0$ = 13.09$\pm$0.14, which was
significantly shorter than all previously determined distances for
47\,Tuc. Using the new value for the hydrogen layer mass they also
provided a new white dwarf distance of $(m-M)_V$ = 13.27 to NGC\,6752.
Shortly afterward \cite{Percival02} re-determined the distance to
47\,Tuc from main-sequence fitting, noting possible calibration
problems with previous studies. They found a best-fit value of
$(m-M)_0 = 13.25^{+0.06}_{-0.07}$, which is in good agreement with the
distance derived from the red clump $(m-M)_0 = 13.31\pm0.05$ and
(within the mutual error bars) with the white dwarf distance derived
by \cite{Zoccali01}.

In 2004 \cite{Moehler04} used FORS1 spectroscopy of white dwarfs in
NGC\,6397 and NGC\,6752 to estimate a distance modulus to NGC\,6397.
The average gravity obtained from multi-colour WFPC2 photometry for
the white dwarfs in NGC\,6752 (assuming a fixed distance modulus) was
used to determine effective temperatures from the spectra for the
white dwarfs in NGC\,6397. Using the effective temperatures determined
that way to estimate the distance of the cluster yielded a true
distance modulus of $(m-M)_0$ = 12.0$\pm$0.1, which is at the short
end of the range of distances derived for NGC\,6397 from main-sequence
fitting ($12.1 \div 12.2$, \cite{Reid98a,Reid98b}).  This result is in
agreement with the one obtained by \cite{Hansen07} ($(m-M)_0$ =
12.02$\pm$0.06) from deep ACS photometry (down to $m_{F814W} =
28$) of NGC\,6397 (see below for the age derived from these data).

Using WFPC2 data \cite{Layden05} determined the distance to M\,5
from both the main sequence ($(m-M)_V = 14.56\pm0.10$, consistent with
previous determinations) and the white dwarf sequence ($(m-M)_V =
14.78\pm0.18$). They ascribe the rather large error of their white
dwarf distance to the low S/N of the observed white dwarfs, the small
baseline in colour ($V-I$), and the mediocre spatial resolution of
their WFPC2 data. These arguments should be considered in further
attempts to determine white dwarf distances to globular clusters.

In a recent investigation \cite{Bonoetal2008} showed that the
relative distances between $\omega$\,Cen and 47\,Tuc based on
different distance indicators (tip of the RGB [TRGB], RR Lyrae,
cluster kinematic) agree within one $\sigma$. However, absolute
kinematic distance moduli are 0.2--0.3\,mag smaller than distances
based on the other methods. The same outcome apply to the distances to
47\,Tuc based on the white dwarf cooling sequence and on the zero age
horizontal branch --- they agree within 0.1\,mag, but they are on average
0.1--0.3\,mag smaller than the distances based on the TRGB and on the RR
Lyrae.

\subsection{Cluster Age Determinations}\label{GCsec:Age}

The white dwarf sequence also provides a possibility to determine the
age of a globular cluster.  However, aside from the observational
difficulties and the uncertainties in the cooling tracks (see
\cite{Chabrier00} for more details) any error in the assumed mass
affects the result. Very deep WFPC2 observations allowed \cite{Richer02}
to detect the white dwarf cooling sequence in M\,4 to
unprecedented depths of $V$$\approx$30.  As a
preliminary result \cite{Hansen02} derived an age of 12.7$\pm$0.7\,Gyr
from the white dwarf luminosity function of M\,4, where the cluster
membership of the stars had been verified from proper motions. This
age is consistent with other independent age estimates, but one should
keep in mind that their error bar does not include errors
due to the uncertainty of the white dwarf mass. Their result had been
questioned by \cite{deMarchi04}, who claimed that the cluster
membership of the white dwarfs cannot be verified down to sufficiently
faint limits to obtain more than a lower limit of the age. In
response to that claim \cite{Richer04} argued that with different
methods for data reduction and analysis different limiting magnitudes
can be reached using the same data.

In a more detailed paper \cite{Hansen04} carefully
studied the influence of the following parameters on the age
determination: distance, reddening, atmospheric models, cooling
models, main sequence lifetime, initial-to-final mass relation, core
composition, atmospheric composition, and binary fraction.

To determine the age they performed a 2-dimensional modeling of the
white dwarf cooling sequence, i.e. they modeled magnitude \emph{and}
colour. They started from a main sequence initial mass function (IMF),
using an initial-to-final mass relation, and then assigned colour and
magnitude based on the white dwarf cooling age. This way they derived
an age of 12.4$^{+1.8}_{-1.5}$\,Gyr (2$\sigma$ limits) for an IMF slope
($dN/dM_{ms} \sim M_{ms}^{-(1+x)}$) of $x=-$0.85. Variations of the
input parameters envelope mass, C/O ratio in the core, initial mass
function, cooling models, atmospheric composition (H-rich
vs. He-rich), atmospheric models, binary fraction, proper motion
cutoff, and distance yielded a global best fit for an age of 12.1\,Gyr
and an IMF slope of $x=-$1.2 with a 2$\sigma$ lower age limit of
10.3\,Gyr, in good agreement with other determinations, but with
substantially larger uncertainties than their preliminary result in
\cite{Hansen02}. IMF slope and age are correlated as larger ages
require that a larger fraction of the white dwarfs has cooled to below
the detection limit, so that a smaller value of $x$ is required to fit
the observed star counts. One should keep in mind that the $x$ values
derived by the fitting procedure are local ones (based on one
region within the globular cluster) and may be influenced by the
effects of mass segregation.

They estimated a current birth rate of white dwarfs of
1.5$\times$10$^{-6}$\,yr$^{-1}$, which is consistent with the birth
rate of horizontal branch stars for that cluster.

In their analysis of ACS observations of M\,4
\cite{Bedinetal2009} concentrate on the detection of the faintest
objects. This was achieved by restricting the searches for those
objects to areas with an as flat as possible background (e.g. far from
bright stars). This way they reach $m_{F606W}$ = 28.95 (50\%
completeness) in about 20\% of the covered area (compared to 26.92 for
the full region). The method is verified with similar data on
NGC\,6397. In the deep areas they reach the end of the white dwarf
cooling sequence in M\,4, from which they derive an age of
11.6$\pm$0.6\,Gyr (internal errors only). This is consistent with the
age of 12.0$\pm$1.4\,Gyr derived from the main sequence turnoff. This
paper as well as \cite{Hansen04} are critically discussed by
\cite{Kaliraietal2009}, who point out various uncertainties not
addressed in the original papers.

According to \cite{Richeretal2008} the peak in the white dwarf
luminosity function is expected to move by about 1\,mag/Gyr for ages
of 12--14\,Gyr, providing a very sensitive age indicator if the white
dwarf cooling sequence is caught completely. 

Using deep ACS photometry (down to $m_{F814W} = 28$) of NGC\,6397
\cite{Hansen07} derived age, reddening, distance, and an
initial-to-final mass relationship from a fit of the complete white
dwarf sequence (magnitude \emph{and} colour, similar to their work on
M\,4 \cite{Hansen04}). Using various evolutionary sequences they
arrived at an age of 11.47$\pm$0.47\,Gyr (95\% confidence limits) in
agreement with, but more precise than, previous determinations from
the main sequence turnoff. All models, however, fail to reproduce the
full extent of the blueward turn at the faint end of the observed
white dwarf cooling sequence. The initial-to-final mass relation which
they derived simultaneously from their data is in good agreement with
the empirical one from \cite{Weidemann00} and also with the white
dwarf mass estimated at the bright end by \cite{Moehler04}.

On the theoretical side, \cite{PradaMoroni07} addressed the
uncertainties affecting cluster age determinations based on the white
dwarf luminosity function. In particular, they found that different
assumptions concerning the conductive opacities might affect the ages
of globular clusters at the 10\% level, while the impact of the C/O
ratio in the core is smaller than 5\%. Interestingly enough, they also
found that the cluster age is only marginally affected by the adopted
initial mass function, since it does not affect the position of the
peak in the luminosity function. On the other hand, the adopted
initial-to-final mass relation affects both the shape and the position
of the peak of the luminosity function. The change causes an
uncertainty on the cluster age of about 8\%. A similar uncertainty on
the cluster age is also caused by the metal abundance, and indeed for
progenitors with metallicities ranging from Z=0.0001 to Z=0.001 and
from Z=0.001 to Z=0.006 the difference is of the order of 10\% on
average (see also \cite{Salaris00} and \cite{PradaMoroni02}).

In a recent investigation \cite{Kowa07} included $Ly_\alpha$ red
wing opacity in the pure hydrogen white dwarf atmosphere models. He
found that this physical ingredient plays a relevant role in the
analysis of cluster white dwarf cooling sequences. The application of
the new models to NGC\,6397 suggests that the white dwarfs at the end of
its cooling sequence are about 160\,K cooler, and in turn that the
cluster is about 0.5\,Gyr older than previously estimated by
\cite{Hansen07}.

\section{Future Perspectives}\label{GCsec:Future}

While several recent photometric and spectroscopic investigations of
cluster white dwarfs significantly improved our knowledge of these
objects we are still facing longstanding problems. The right
instruments to properly address these problems will be future
extremely large telescopes like the European Extremely Large Telescope
(E-ELT\footnote{http://www.eso.org/projects/e-elt/}, \cite{Hook06};
\cite{Bono07}), the Thirty Meter
Telescope\footnote{http://www.tmt.org/}, or the Giant Magellan
Telescope\footnote{http://www.gmto.org/}. The use of multi-conjugate
adaptive optic systems (\cite{Marchetti06}) will provide near-infrared
(NIR) images with a full width at half maximum (FWHM) of the order of
a few hundredths of arc seconds, a field of view (FoV) of a few
arcminutes squared and very high spatial resolution (from 0.005'' to
0.01'', see, e.g., \cite{ELT}).  The same outcome applies to the James
Webb Space Telescope\footnote{http://www.stsci.edu/jwst/}. This
telescope will be equipped with a NIR Camera
(NIRCam\footnote{http://ircamera.as.arizona.edu/nircam/}) covering the
wavelength range from 0.6 to 5\,$\mu$m with a FoV of
$2.2\times4.4\Box$' for simultaneous observations in two wavelength
ranges.  The short wavelength ($\lambda \le$\,2.4\,$\mu$m) channels
have a spatial sampling of $0.03$''/px, while the long ones have
$0.06$''/px with Nyquist sampling at 2 and 4\,$\mu$m
(\cite{Rie05}). In addition the NIR spectrograph
(NIRspec\footnote{http://sci.esa.int/science-e/www/object/index.cfm?fobjectid=456940})
covers the same wavelength region at low spectral resolution
(R$\approx$3000).  The key advantage of this choice is the possibility
to collect spectra of faint targets located close to bright sources
(\cite{Lobb06}).  This is a very typical situation in the crowded
central regions of globular clusters.  There are several fields where
these observational equipments might play a crucial role.

{\em i)} \emph{Variable Stars} -- The 30m--40m class telescopes will
provide the unique opportunity to collect multi-band photometry of
globular cluster white dwarfs with an unprecedented time
resolution. This means that we should be able to investigate {\em in
situ} the topology of the instability strips located along the cooling
sequence of cluster white dwarfs (\cite{Kim06}). Variable white dwarfs
are multi-periodic non-radial oscillators with periods ranging from a
few tenths to roughly 1000 seconds (\cite{silvo07}) and present small
luminosity amplitudes. Starting at the hot end of the white dwarf
cooling sequence we first find the instability strip of hot
(80,000\,K\,$\le$\,T$_{\rm eff}$\,$\le$\,170,000\,K) either pre-white
dwarfs or central stars of planetary nebulae. They are called DOV or
GW\,Vir stars and the prototype is PG1159$-$035. The instability strip
of DBV stars is located at cooler effective temperatures
(22,000\,K\,$\le$\,T$_{\rm eff}$\,$\le$\,28,000\,K). Eventually, we
approach the instability strip of the DAV or ZZ Ceti variables at even
cooler effective temperatures (11,000\,K $\le {\rm T}_{\rm eff} \le$
12,500\,K). Note that the current knowledge on variable white dwarfs
is based on local objects only, which are affected by uncertainties in
the progenitor mix, the distance, the reddening, and the effective
temperature. Cluster variable white dwarfs will provide firm
constraints on the driving mechanisms and the possible occurrence of
static stars inside the various instability strips
(\cite{Silvotti06}).\\ Moreover, the detection of cluster DBVs might
provide fundamental constraints on plasmon neutrinos. The neutrino
produced in the decay of plasmons cannot be observed in physics
laboratories (\cite{ito96}). However, current theoretical predictions
indicate that half of the luminosity of a hot white dwarf (T$_{\rm
eff}\approx$\,25,000\,K) is carried away by neutrino emission
(\cite{obrien00}; \cite{Kim08}). This means that accurate measurements
of period changes over time for DBV stars can constrain plasmon
neutrino rates (\cite{Winget04}). The great advantage of using cluster
DBVs is the possibility to compare the plasmon neutrino rates based on
pulsation observables with those based on evolutionary observables
(white dwarf luminosity function).  Current luminosity functions of
white dwarfs in globular clusters do not provide firm constraints
on this mechanism as the critical temperature range of the cooling
sequence is too poorly populated.\\ Cluster white dwarf variables can
also play an important role for even more fundamental physics. It has
been suggested (\cite{Alc88}) that compact objects, such as neutron
stars, might consist of Strange Quark Matter ({\bf SQM}). SQM is a
particular form of quark matter with a strangeness per baryon value of
about $-1$ (\cite{Bod71}; \cite{Wit84}).  Theoretical predictions
indicate that SQM may also exist in the cores of white dwarfs
(\cite{Gle95}; \cite{Var04}).  Two different observational approaches
have been suggested to constrain the existence of exotic white dwarfs
with SQM cores and envelopes of normal matter.  {\em i)} The
mass-radius relation. However, the difference between normal and
exotic white dwarfs is too small to be observable (see Fig.~4 in
\cite{Pan00}).  {\em ii)} Gravity modes. Pulsations are very sensitive
to density profiles and \cite{Ben05}, \cite{Ben06a}, and \cite{Ben06b}
demonstrated that white dwarf models with SQM cores show a completely
new resonant cavity for gravity modes.  This causes the period spacing
between consecutive modes in exotic white dwarfs to be shorter than
one second, while in normal white dwarfs it is of the order of several
tens of seconds. Moreover, the number of modes inside the different
clusters of periods strongly depends on the size of the SQM core. The
observational signature of this effect is that the Fourier spectrum of
exotic white dwarfs should show peaks that are characterized by an
intrinsic width, while those of normal white dwarfs become sharper
with increasing time coverage.  Cluster white dwarf variables will
provide the large homogeneous samples required to constrain the
possible occurrence of these phenomena.\\ Longer time series data
covering several hours will also provide fundamental constraints on
the number of cataclysmic variables in globular clusters as well as on
the frequency of dwarf novae.
   
{\em ii)} \emph{Metals and IR Excess} -- White dwarfs have metal-poor
atmospheres. Whatever the metal abundance of the progenitor, heavy
elements in their atmosphere can only survive for a short period
($\approx 10^7$ yr). This empirical evidence is supported by the plain
physical argument that radiative levitation is stronger than
gravitational settling only at the hot end of the white dwarf cooling
sequence (T$_{\rm eff}$\,$\ge$\,20,000\,K, \cite{Cha95}, and
references therein). Cooler white dwarfs are characterized by
diffusion timescales that are several orders of magnitudes shorter
than the cooling lifetime.  Therefore, the occurrence of metal lines
in the photospheres of cool white dwarfs must be due either to the
inward growth of the convective region that dredges-up carbon
(\cite{Far08}) or to accretion of interstellar matter (\cite{Koe05}).
Detailed theoretical calculations indicate that the diffusion
timescale of Ca, Mg, and Fe in ``metal-rich'' DA (DAZ) white dwarfs
are 3--4 orders of magnitude shorter than for ``metal-rich'' non-DA
(DZ) white dwarfs ($10^6$\,yr, \cite{Koe06}). This means that DAZ
white dwarfs provide a unique opportunity to measure the actual
accretion rate and to constrain small scale variations of the
composition of the interstellar medium by comparing the composition of
the white dwarfs (dominated by accretion while moving through the
interstellar medium) to the local properties of the interstellar
medium. The detection of cluster DAZ white dwarfs will provide the
opportunity to measure the efficiency of pollution and in particular
to probe the extent and composition of intra-cluster material. This
questions becomes even more intriguing if one accounts for the fact
that one scenario to explain the nature of the contaminating elements
envisages the tidal disruption of an asteroid and an ensuing infrared
excess (\cite{Jur03}). This working hypothesis is supported by the
discovery of several systems in the field showing both metal lines and
a well defined infrared excess during the last few years
(\cite{Jur07a}; \cite{vonh07}).\\ Moreover, \cite{Sig03} and
\cite{Richer03} suggest that the millisecond pulsar PSR\,B1620$-$26 in
the globular cluster M\,4 is a triple system including a white dwarf
and a third body with a mass value of the order of a few Jupiter
masses. The detection of a planet in a globular cluster of
intermediate metallicity is relevant not only for constraining
planetary formation mechanisms but also to investigate their survival
rate in old dense stellar systems. The possible occurrence of planets
in globular clusters might also have an impact on the evolutionary
properties of cluster stars. It has been suggested by \cite{Sok07}
that the presence of planets could affect the evolution of the parent
stars, and in turn the morphology of the horizontal branch and the
presence of low-mass white dwarfs.\\ The occurrence of infrared excess
among white dwarfs is typically explained as the presence either of
dust (disks, circumstellar envelope) or of a low-luminosity
companion. Thanks to near-
(2MASS\footnote{http://irsa.ipac.caltech.edu/Missions/2mass.html}) and
mid-IR
(Spitzer\footnote{http://sha.ipac.caltech.edu/applications/Spitzer/SHA/})
photometric surveys a sizable fraction of white dwarfs hosting both
stellar and substellar (brown dwarfs) companions have been identified
(\cite{Wac03}; \cite{Jur07b}; \cite{Deb07}, and references
therein). The detection of cluster white dwarfs that show a well
defined NIR excess can provide useful constraints on the current
fraction of binary stars, but also on the evolutionary properties of
brown dwarfs in metal-poor regimes. The current and the next
generation of NIR adaptive optics systems at the 10m class
telescopes will allow us to tackle this important experiment. The quality
of the images (FWHM$\le$0.1'' in the K-band) and the field of view
(1$\Box$') provide the opportunity to perform accurate photometry for
large samples in crowded cluster regions down to limiting magnitudes
of $K\approx$21.5--22 with exposure times of less than one hour
(\cite{Bono08}).

{\em iii)} \emph{Spectral Classification} -- Thanks to its superb
spatial resolution and its larger collecting area the 30m--40m class
telescopes will allow us to observe spectra of much fainter white
dwarfs than today. This will be essential to answer the question of
the ratio DA:non-DA in globular clusters. 

{\em iv)} \emph{Missing Physical Ingredients} -- The comparison
between theory and observations is a fundamental step in constraining
the plausibility and the accuracy of the physical assumptions adopted
to construct stellar models. With respect to globular cluster white
dwarfs we are still facing several open problems.  The mismatch
between the observed blueward turn of the white dwarf cooling sequence in
NGC\,6397 and synthetic colour-magnitude diagrams indicates that theory
shows a smoother transition to bluer colours than is observed.
A plausible culprit of this discrepancy might be an
underestimate of the CIA from molecular
hydrogen.  This working hypothesis is supported by recent theoretical
investigations suggesting that new DA white dwarfs atmosphere models accounting
for the $Ly_\alpha$ red wing opacity are systematically redder that
the old ones (\cite{Kowa07}).  One of the two anonymous referees
properly noted that quantitative constraints on the currently adopted
CIA have also significant impact on the predictions of planetary
spectra.\\   
Circumstantial empirical evidence indicates that globular clusters with
extremely hot horizontal branch stars (hot helium flashers,
cf. Sec.~\ref{GCsec:Helium}) present an excess of white dwarfs when
compared with main sequence turnoff stars. It has been suggested that
a fraction of He-core white dwarfs might account for current star
counts. However, the number of globular clusters for which
sufficiently accurate and deep data are available is quite limited and
we still also lack firm constraints on the accuracy of cooling lifetimes in
the high temperature range (\cite{Cala08}).\\

Summarizing the previous sections we feel quite confident that the
  coming years will provide ample material for new reviews on white
  dwarfs in globular clusters -- and we look forward to reading them!

\acknowledgement It is a pleasure to thank C.E. Corsi, G. Prada Moroni
and M. Salaris for several useful discussions on white dwarfs, and H.
Kuntschner for information on the instruments of the James Webb Space
Telescope. This work was partially supported by ASI (P.I.: F. Ferraro)
and by PRIN-MIUR~2009 (P.I. R.~Gratton). One of us, G.B., acknowledges
support from the ESO Visitor program.  Extensive use has been made of
both the ADS and the arXiv preprint archive. We also thank two
anonymous referees for their suggestions and comments that helped to
improve the quality of the manuscript.

%
%

%
%

\end{document}